\documentclass[10pt,letterpaper]{article}
\usepackage{opex3}
\usepackage{color}
\usepackage[latin1]{inputenc}
\usepackage[english]{babel}
\usepackage{graphicx}
\usepackage[caption=false]{subfig}
\usepackage{amsmath}
\usepackage{float}
\usepackage{cite}
\usepackage{bbold}

\newcommand{\mat}[1]{\ensuremath{\mathrm{#1}}}
\newcommand{\gmat}[1]{\ensuremath{\boldsymbol{#1}}}

\graphicspath{{Figures/}}

\begin{document}

%%%%%%%%%%%%%%%%%% title page information %%%%%%%%%%%%%%%%%%
\title{Single-quadrature continuous-variable quantum key distribution}

\author{Tobias Gehring,$^1$ Christian S. Jacobsen,$^1$ and Ulrik L. Andersen$^{1*}$}
\address{$^1$ Department of Physics, Technical University of Denmark, Fysikvej, 2800 Kongens Lyngby, Denmark}
\email{$^*$ ulrik.andersen@fysik.dtu.dk}

\begin{abstract}
Most continuous-variable quantum key distribution schemes are based on the Gaussian modulation of coherent states followed by continuous quadrature detection using homodyne detectors.
In all previous schemes, the Gaussian modulation has been carried out in conjugate quadratures thus requiring two independent modulators for their implementations.
Here, we propose and experimentally test a largely simplified scheme in which the Gaussian modulation is performed in a single quadrature.
The scheme is shown to be asymptotically secure against collective attacks, and considers asymmetric preparation and excess noise. A single-quadrature modulation approach renders the need for a costly amplitude modulator unnecessary, and thus facilitates commercialization of continuous-variable quantum key distribution. 
\end{abstract}


%%%%%%%%%%%%%%%%%%%%%%% References %%%%%%%%%%%%%%%%%%%%%%%%%

%%%%%%%%%%%%%%%%%%%%%%%%%%  body  %%%%%%%%%%%%%%%%%%%%%%%%%%

\section{Introduction}

The quantum informational primitive of quantum key distribution (QKD) allows two parties (Alice and Bob) to distill a secret key using an untrusted quantum channel and an authenticated classical channel \cite{Scarani2009}.
Various forms of QKD have been proposed and experimentally realized in laboratories and under real-life field conditions \cite{Lo2014,Diamanti2015}, and they can be roughly divided into two different categories which depend on the actual measurement strategy at Bob's station: Measurement of a discrete variable (DV) -- the photon number -- carried out by photon counters or measurement of a continuous variable (CV) \cite{Weedbrook2012} -- the field quadratures -- performed by a homodyne detector. 
They are referred to as DVQKD and CVQKD, respectively.

While DVQKD is the most matured scheme in terms of security proofs~\cite{Lim2014,Curty2014}, secure communication distance \cite{Korzh2015} and real-life field tests \cite{Peev2009,Sasaki2011}, the scheme of CVQKD is rapidly becoming a serious competitor due to recent promising developments.
This includes the recent advances in deriving composable security proofs \cite{Furrer2012,Furrer2014,Leverrier2015}, the development of very efficient post-processing algorithms for the distillation of a secret shared key from the raw key \cite{Jouguet2014a,Jouguet2014b,Gehring2015}, the implementations of long-distance QKD \cite{Jouguet2013} and the recent developments of more advanced protocols such as measurement-device-independent CVQKD \cite{Pirandola2015} and squeezed states QKD~\cite{Gehring2015,Madsen2012}.
Moreover, most protocols of CVQKD benefit from the fact that the associated technology required for real-world integration is based on standard telecommunication components.  

Coherent state based CVQKD protocols can be distinguished by the different input alphabets.
Previous proposals have included discrete \cite{Ralph1999,Hillery2000,Reid2000} as well as continuous \cite{Grosshans2002,Grosshans2003} modulation patterns of Gaussian states in phase space.
Most of these schemes are based on the modulation of states in conjugate bases, similarly to the famous BB84 protocol \cite{Bennett84} where a polarization eigenstate is modulated in conjugate bases.
In BB84, the swapping between conjugate bases is an absolute necessity for establishing security, but this is not the case for the modulation of Gaussian states.
In this case, the modulated states are not eigenstates of the modulation basis (in contrast to BB84) but non-orthogonal, and thus the basic non-orthogonality requirement for secure communication is fulfilled even in a single basis \cite{Bennett1992,Lorenz2006}.
Single-quadrature modulation has been considered for a two-state protocol with coherent states \cite{Lorenz2006,Zhao2009} and for the continuous modulation of squeezed states \cite{Jacobsen2014} followed by homodyne detection. 

In this Letter we develop a protocol using modulated coherent states in a single quadrature. 
Our security analysis includes excess and preparation noise, and we present a proof-of-principle experimental implementation of the protocol using both homodyne and heterodyne detection.
Single-quadrature CVQKD is highly relevant as it allows for a simplification of the required technology at the sending station and, thus, leads to an important reduction in the cost.
For a dual-quadrature Gaussian modulation scheme, an amplitude as well as a phase modulator is needed while the single-quadrature scheme can be implemented with a single modulator, e.g.\ leaving out the amplitude modulator.
This constitutes a significant reduction in complexity and cost as an amplitude modulator is based on an interferometric configuration (that must be stabilized) and is very expensive in terms of optical power consumption.

In preparation of this manuscript we discovered that the idea of a single quadrature CVQKD protocol has been independently developed by Usenko \textit{et al.}~\cite{Usenko2015}, where the implications of an asymmetric channel are considered. 
We finalized our security analysis with inspiration from this work.

\section{Theory}

\begin{figure}[b]
    \centering
    \includegraphics[width=8.5cm]{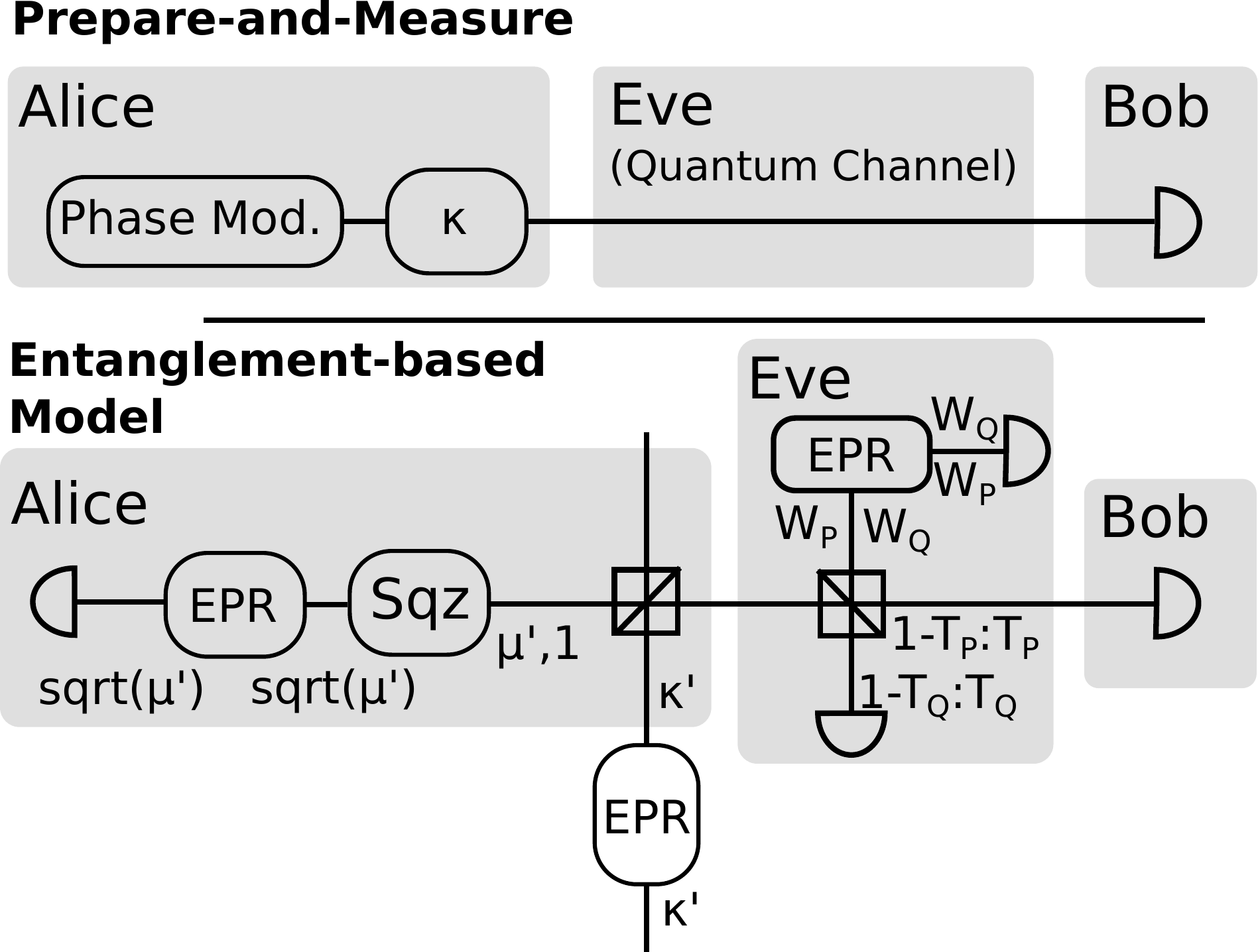}
    \caption{Prepare-and-measure protocol based on modulating a single quadrature (phase) and equivalent entanglement-based model used to calculate the secret key rate. EPR: Einstein-Podolsky-Rose source, Sqz: Squeezing operation.}
    \label{fig:scheme}
\end{figure}
We consider the Prepare-and-Measure CVQKD protocol illustrated in Fig.~\ref{fig:scheme}.
Random numbers drawn from a one-dimensional Gaussian alphabet are used to modulate the phase quadrature, $P$, at Alice's station, thereby preparing independent coherent states along a line in phase space.
This is in stark contrast to previous protocols where a two-dimensional Gaussian distribution is used to drive phase and amplitude quadratures.
The prepared states are sent to Bob who performs a coherent detection of the quadratures, either using a heterodyne detector or a homodyne detector that switches between an amplitude and a phase quadrature measurement.
The list of data obtained by Bob (when he measured the P quadrature) is correlated with the list of Alice, and this correlation is subsequently used to generate a secret key by means of error reconciliation and privacy amplification.  

The security of the scheme can be addressed by using the theoretical equivalence between the prepare-and-measure scheme and an entanglement-based scheme~\cite{Grosshans2003a,GarciaPatron2007}, see Fig.~\ref{fig:scheme}. 
In the entanglement-based scheme, Alice prepares an Einstein-Podolsky-Rosen (EPR) state of variance $\sqrt{\mu'}$, keeps one mode of the EPR state to herself and sends the other mode to Bob.
If Alice performs a heterodyne measurement on her mode (that is projecting it onto a coherent state), the remaining mode will form a 2D Gaussian distribution of coherent states while a homodyne measurement at Alice will form a 1D Gaussian distribution of squeezed states.
A 1D distribution of coherent states can thus be realized by performing a local squeezing operation with a squeezing parameter of $\xi = \log(\sqrt[4]{\mu'})$ onto the second mode before it is sent to Bob. 
Assuming that the covariance matrix of the EPR state has the symmetric form
\begin{equation}
\gmat{\Gamma_{AB}'} =
\begin{bmatrix}
\mathbb{1} \sqrt{\mu'} & \mathbb{Z} \sqrt{\mu' - 1}  \\
\mathbb{Z} \sqrt{\mu' - 1} & \mathbb{1} \sqrt{\mu'} 
\end{bmatrix}
\end{equation}
with
\begin{equation}
\mathbb{1} =
\begin{bmatrix}
1 & 0 \\
0 & 1
\end{bmatrix}
\qquad , \qquad
\mathbb{Z} =
\begin{bmatrix}
1 & 0 \\
0 & -1
\end{bmatrix}
\end{equation}
then the local squeezing transforms it to the matrix
\begin{equation}
\gmat{\Gamma_{AB}} =
\begin{bmatrix}
\sqrt{\mu'} & 0 & \dfrac{\sqrt{\mu' - 1}}{\sqrt[4]{\mu'}} & 0 \\
0 & \sqrt{\mu'} & 0 & -\sqrt[4]{\mu'} \sqrt{\mu' - 1} \\
\dfrac{\sqrt{\mu' - 1}}{\sqrt[4]{\mu'}} & 0 & 1 & 0 \\
0 & -\sqrt[4]{\mu'} \sqrt{\mu' - 1} & 0 & \mu'
\end{bmatrix}\quad .
\end{equation}

To assess the security of the proposed scheme against collective attacks, we consider a generalized Gaussian attack, the asymmetric entangling cloner attack, which is the most powerful attack that can be performed on the quantum channel, in the limit of infinite exchanges, where Gaussian extremality holds true~\cite{Weedbrook2012,Grosshans2002}.
Such an attack can be executed by a local unitary equivalent to an asymmetric beam splitter with one mode of an EPR state controlled by Eve in the secondary input port. She also has control over the beam splitting ratio and the degree of asymmetricity, and is limited only by the laws of quantum mechanics.
Eve keeps the other mode of her EPR state in a quantum memory and interferes the first mode with the coherent states sent by Alice. The interfered mode is saved in another quantum memory.
Eve's information gain in the limit of infinitely many uses of the channel is upper bounded by the Holevo quantity \cite{Weedbrook2012}.

In this asymptotic limit, the asymptotic equipartition property applies and both the Shannon and the von Neumann entropy are well defined quantities \cite{Nielsen2000,Cover2006}.
Finite key sizes have a subtle impact on the security proofs \cite{Renner2005,Leverrier2015} which will not be considered in this paper.
As a first approximation, neglecting these issues, one arrives at the following bound on the secret key rate~\cite{Scarani2009,Weedbrook2012},
\begin{equation} \label{eq:keyrate}
R = \beta I(A:B) - \chi(E:X)\ ,
\end{equation}
where $\beta$ is the reconciliation efficiency and $I(A:B)$ is the classical mutual information between Alice and Bob expressed through the Shannon entropy of the corresponding classical stochastic variables of the measurements \cite{Cover2006}.
$\chi(E:X)$ is the Holevo quantity \cite{Nielsen2000} between Alice (Bob) and Eve for direct (reverse) reconciliation, which can be expressed by the von Neumann entropy $S(\rho) = S(\gmat{\Gamma})$ of the quantum state $\rho$, $\chi(E:X) = S(E) - S(E|X)$~\cite{Weedbrook2012}. For Gaussian states with zero mean, $\rho$ can be completely described by its covariance matrix $\gmat{\Gamma}$.

The von Neumann entropy for a Gaussian state $\rho$ is given by~\cite{Weedbrook2012}
\begin{equation} \label{eq:vonNeumannGamma}
S(\gmat{\Gamma}) = \sum_i g(\nu_i)\ ,
\end{equation}
where
\begin{equation} \label{eq:BosonInfo}
g(x) = \dfrac{x+1}{2} \log_2 \left(\dfrac{x+1}{2} \right) -  \dfrac{x-1}{2} \log_2 \left(\dfrac{x-1}{2} \right)\ ,
\end{equation}
and $\nu_i$ is the $i$'th value in the symplectic spectrum of $\gmat{\Gamma}$, and the function $g(x)$ is normalized in units of vacuum noise. 
The symplectic spectrum is calculated by finding the absolute eigenvalues of the matrix $i \gmat{\Omega} \gmat{\Gamma}$, where
\begin{equation}
\gmat{\Omega} = \bigoplus_{k=1}^N \gmat{\omega}
\qquad , \qquad
\gmat{\omega} = 
\begin{bmatrix}
0 & 1 \\
-1 & 0
\end{bmatrix}\ ,
\end{equation}
with $N$ being the number of modes described by the state $\rho$. 

We include trusted preparation noise in our model to account for its presence in the experimental implementation~\cite{Filip2008,Usenko2010,Weedbrook2010,Weedbrook2012a}.
Theoretically, this is simulated by assuming that the environment prepares an EPR state, one mode of which is interfered with the signal mode on a beam splitter with splitting ratio $\eta \approx 1$ before it enters the quantum channel. 
In the limit of no preparation noise, this beam splitter will of course substitute part of the EPR state with vacuum noise, but an appropriate redefinition of the system parameters makes this error insignificant. More importantly, this approximation does not overestimate the security of the protocol, since the dilution only reduces the correlations between Alice and Bob. We get the following covariance matrix,

\begin{equation}
\gmat{\Gamma_{A \kappa'}} = 
\begin{bmatrix}
\mat{A'} & \mat{C'} \\
\mat{{C'}}^T & \mat{K'}
\end{bmatrix},
\end{equation}

with the submatrices 

\begin{equation}
\mat{A'} = 
\begin{bmatrix}
\sqrt{\mu'} & 0 & \dfrac{\sqrt{\eta} (\mu' - 1)}{\sqrt[4]{\mu'}} & 0 \\
0 & \sqrt{\mu'} & 0 & -\sqrt{\eta} (\mu' - 1)\sqrt[4]{\mu'} \\
\dfrac{\sqrt{\eta} (\mu' - 1)}{\sqrt[4]{\mu'}} & 0 & \eta + (1-\eta) e^{-2r} \kappa' & 0 \\
0 & -\sqrt{\eta} (\mu' - 1)\sqrt[4]{\mu'} & 0 & \eta \mu' + (1-\eta)e^{2r} \kappa'
\end{bmatrix}
\end{equation}

\begin{equation}
\mat{K'} = 
\begin{bmatrix}
1 - \eta + \eta e^{-2r} \kappa' & 0 & \sqrt{\eta({\kappa'}^2 - 1)} e^{-r} & 0 \\
0 & (1-\eta) \mu' + \eta e^{2r} \kappa' & 0 & -\sqrt{\eta({\kappa'}^2 - 1)} e^{r} \\
\sqrt{\eta({\kappa'}^2 - 1)} e^{-r} & 0 & \kappa' & 0 \\
0 & -\sqrt{\eta({\kappa'}^2 - 1)} e^{r} & 0 & \kappa'
\end{bmatrix}
\end{equation}

\begin{equation}
\mat{C'} = 
\begin{bmatrix}
- \dfrac{\tilde{\eta} \sqrt{\mu'-1}}{\sqrt[4]{\mu'}} & 0 & 0 & 0 \\
0 & \tilde{\eta} \sqrt{\mu'-1} \sqrt[4]{\mu'} & 0 & 0 \\
\tilde{\eta} \sqrt{\eta} (e^{-2r} \kappa' -1) & 0 & \tilde{\eta} \sqrt{{\kappa'}^2 -1} e^{-r} & 0 \\
0 & \tilde{\eta} \sqrt{\eta}(e^{2r} \kappa' - \mu') & 0 & \tilde{\eta} \sqrt{{\kappa'}^2 -1} e^{r}
\end{bmatrix} \ ,
\end{equation}

where $\tilde{\eta} = \sqrt{1-\eta}$. $\mu'$ and $\kappa'$ are the parameters before the redefinition, and the squeezing parameter $r$ is used to determine the asymmetry of the preparation noise. The dummy parameters are related to the actual protocol parameters by the following equations

\begin{align}
& \eta \mu' + (1-\eta)e^{2r} \kappa' = \mu + \kappa_P\ ,
\\
& \eta + (1-\eta)e^{-2r} \kappa' = 1 + \kappa_Q\ ,
\\
& \dfrac{\eta(\mu' - 1)}{\sqrt{\mu'}} = \dfrac{\mu - 1}{\sqrt{\mu}}\ ,
\end{align}

which have the following solutions

\begin{align}
& \kappa' = \dfrac{\mu + \kappa_P - \eta \mu'}{(1-\eta)} e^{-2r}\ ,
\\
& r = \ln \left(\dfrac{\mu + \kappa_P - \eta \mu'}{1 + \kappa_Q - \eta} \right)/4\ ,
\\
& {\mu'}^2 = \dfrac{\Delta}{6 \eta} + \dfrac{2 \eta}{\Delta}\ ,
\end{align}

where the determinant $\Delta$ is given by
\begin{equation}
\Delta = \sqrt[3]{(12 \sqrt{3} \sqrt{27 \mu^3 - 54 \mu^2 - 4 \eta^2 + 27 \mu} + 108 \sqrt{\mu^3} - 108 \sqrt{\mu})\eta^2}\ .
\end{equation}

After this redefinition we assume the preparation noise to be added to our signal beam to have the variance $\kappa_P$ in the phase quadrature and $\kappa_Q$ in the amplitude quadrature, while the signal variance in the phase quadrature is $\mu - 1$.
The environmental modes are not accessible to the eavesdropper and so the noise is trusted. Letting the environment prepare an EPR state in this manner conveniently leaves the global state pure.

We propagate $\gmat{\Gamma_{A \kappa'}}$ through the quantum channel with asymmetric loss parameters $T_Q$ and $T_P$ for the amplitude and phase quadratures, respectively, also interfering with Eve's noisy EPR state with variances $W_Q$ and $W_P$, respectively.

The covariance matrix of the global state following the quantum channel is quite large, so only the most important submatrix will be presented here.
This is the matrix that represents the two-mode state shared by Alice and Bob.

\begin{equation} \label{eq:AliceBobPostChannel}
\mat{A} = 
\begin{bmatrix}
\sqrt{\mu'} & 0 & \dfrac{\sqrt{T_Q \eta V_S'}}{\sqrt[4]{\mu'}} & 0 \\
0 & \sqrt{\mu'} & 0 & -\sqrt{T_P \eta V_S'}\sqrt[4]{\mu'} \\
\dfrac{\sqrt{T_Q \eta V_S'}}{\sqrt[4]{\mu'}} & 0 & T_Q \kappa_Q + (1-T_Q) W_Q & 0 \\
0 & -\sqrt{T_P \eta V_S'}\sqrt[4]{\mu'} & 0 & T_P(\mu + \kappa_P) + (1-T_P) W_P
\end{bmatrix} \ ,
\end{equation}

with $V_S' = \mu' - 1$. Firstly, the classical mutual information between Alice and Bob is, after some algebra, found to be
\begin{equation}
I_{\text{Homo}}(A:B) = \dfrac{1}{2} \log_2 \left(\dfrac{(1 - T_P)W_P + T_P (\mu + \kappa_P)}{(1 - T_P)W_P + T_P (\eta + (1-\eta) \kappa' _P)} \right)\ .
\end{equation}

In the limit of $\kappa_P = 0$ and $\eta = 1$ this is similar to the expression one would expect from a dual quadrature protocol with half the alphabet. For heterodyne detection the expression is quite similar,
\begin{equation}
I_{\text{Hete}}(A:B) = \dfrac{1}{2} \log_2 \left(\dfrac{(1 - T_P)W_P + T_P (\mu + \kappa_P) + 1}{(1 - T_P)W_P + T_P (\eta + (1-\eta) \kappa' _P) + 1} \right)\ ,
\end{equation}
with the addition of a unit of vacuum.
The relevant transmission in both cases is the one in the $P$-quadrature, since this is the encoded quadrature.
Similarly, only the excess noise in the $P$ quadrature degrades the mutual information.
However, the excess noise in the conjugate quadrature has a profound impact on the security, which will be elaborated on later.

The expression for the Holevo bound is significantly more complicated.
We begin by considering the quantity $S(E)$, which denotes the von Neumann entropy of Eve's EPR state.
If we assume that the eavesdropper is able to purify the global state, it is easy to show that $S(E) = S(AB\kappa)$, where $S(AB\kappa)$ is the von Neumann entropy of the total multimode system of Alice, Bob and the EPR state which injects the preparation noise.
Purification is a useful technique in this case because it allows one to estimate the information gain of the eavesdropper without having access to her measurements.
Because of the added complication of preparation noise, there is no (short) closed expression for the von Neumann entropy of this state. The symplectic spectrum of $\gmat{\Gamma_{A B \kappa}}$, which represents this global state is easily calculated numerically, for a given choice of channel and input parameters.
Similarly, purification allows the conditional von Neumann entropies to be written as $S(E|B) = S(A\kappa|B)$.
Again, since the symplectic eigenvalues, which are used to calculate the von Neumann entropies, do not permit short analytical expressions, they are calculated numerically and inserted into Eqs.~\eqref{eq:keyrate}, \eqref{eq:vonNeumannGamma} and \eqref{eq:BosonInfo} to calculate the secret key rate.

Since preparation noise has been incorporated into the security analysis, it makes sense to also investigate the case of direct, rather than reverse, reconciliation since it is known from dual quadrature protocols that it offers an advantage in cases of extremely large preparation noise \cite{Weedbrook2010, Weedbrook2012a, Jacobsen2015}. 
This is done by considering the conditional entropy $S(E|A) = S(B\kappa|A)$ rather than $S(E|B) = S(A\kappa|B)$, but otherwise offers no additional computational complications.

\begin{figure}[ht]
    \begin{center}
    \subfloat{\includegraphics[width=0.49\textwidth]{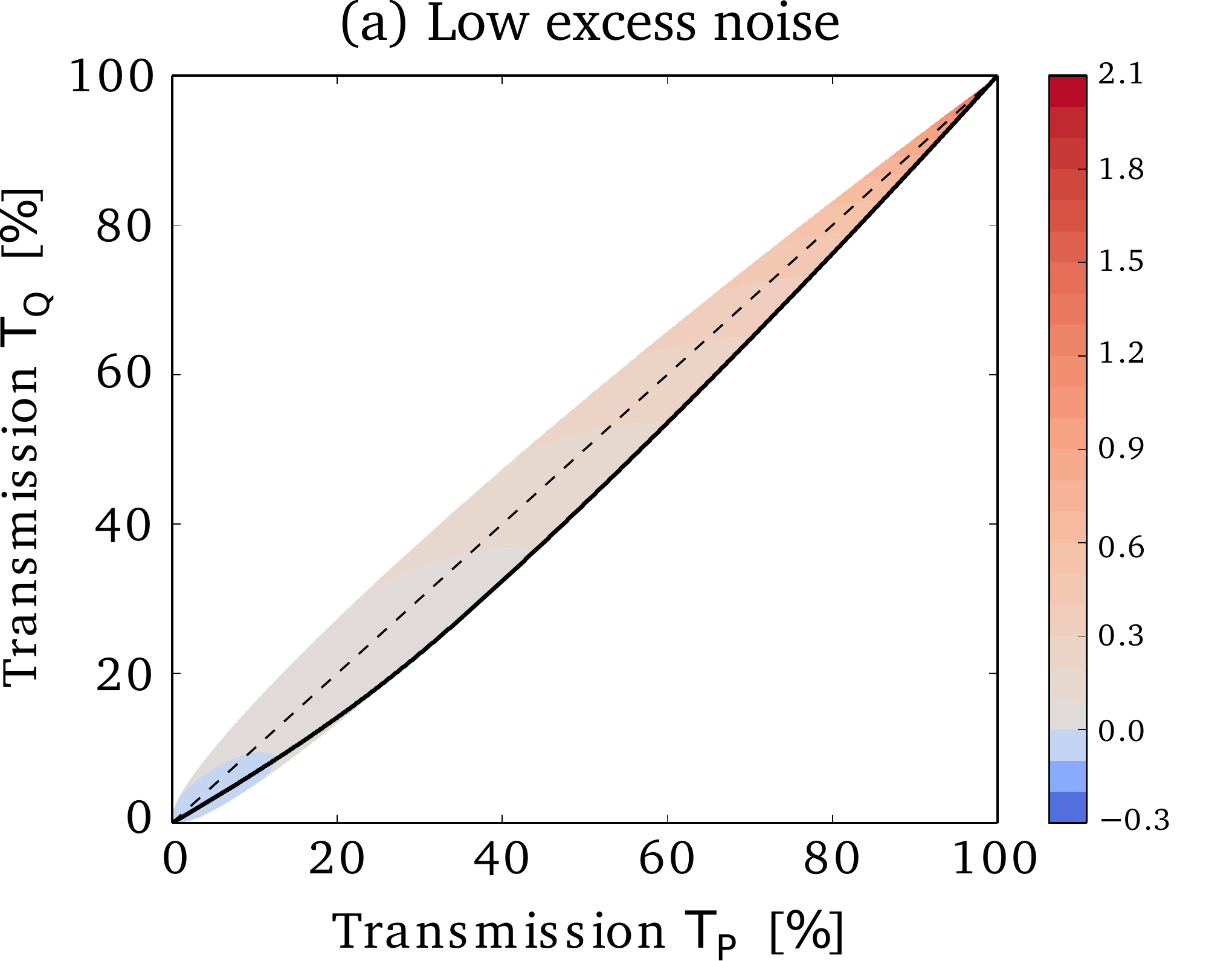}\label{fig:TPTQgridWlow}} 
    ~
    \subfloat{\includegraphics[width=0.49\textwidth]{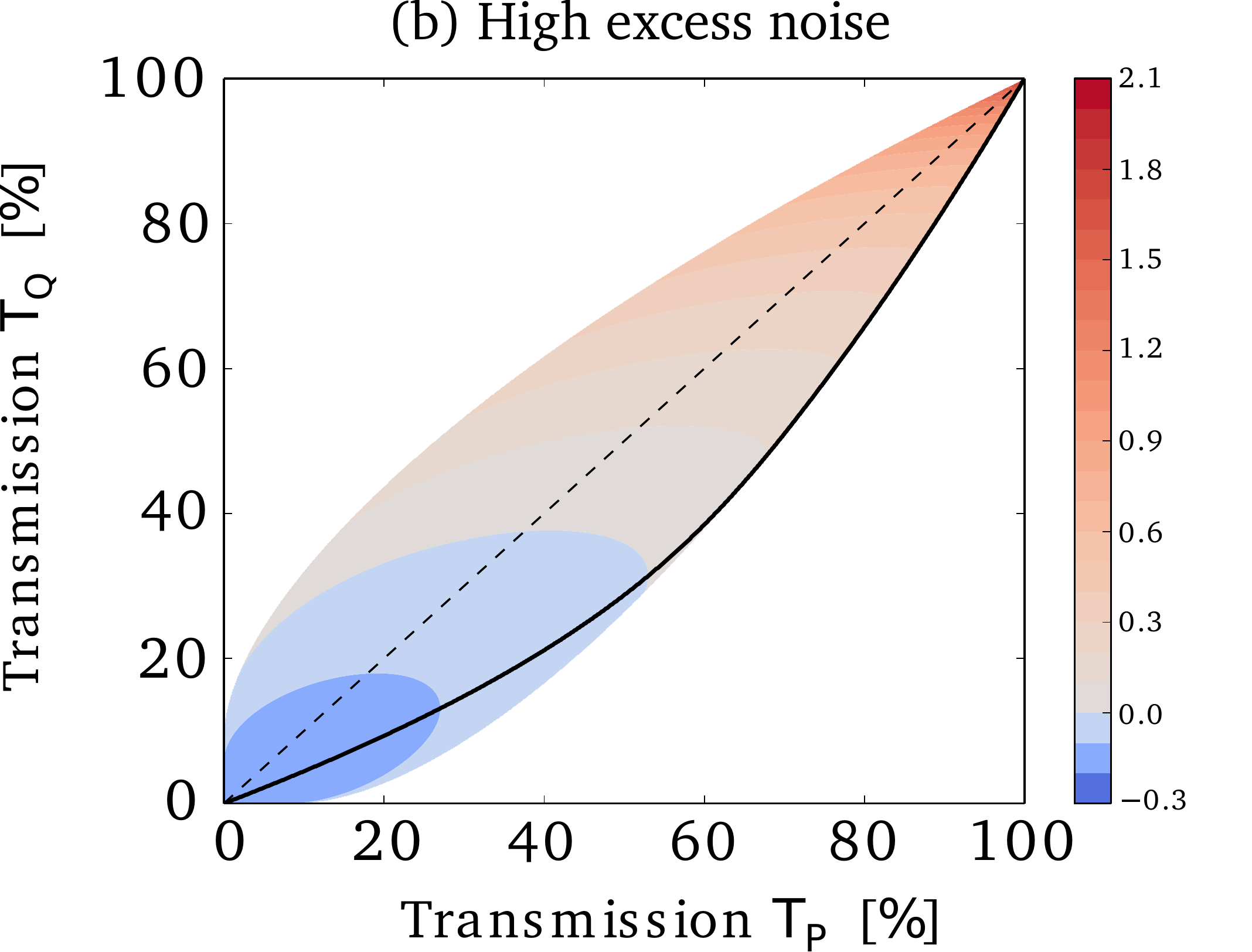}\label{fig:TPTQgridWhigh}}
    \end{center}
\caption{Theory plots of the region of positive secret key rates using single-quadrature modulation with a heterodyne detection strategy in terms of the asymmetric channel loss. The dashed black line is the symmetric value $T_P = T_Q$, and the black solid line is the choice of $T_Q$ that minimizes the rate. Plot (a) has excess noise of $W_P = W_Q = 1.005$ SNU , while for (b) $W_P = W_Q = 1.05$ SNU. $\mu = 31$ SNU, $\beta = 1$ and $\kappa_Q = \kappa_P = 0$.}
    \label{fig:TPTQgrid}
\end{figure}

A security proof for a dual quadrature protocol usually assumes symmetry between the quadratures, but for the presented protocol this is naturally not the case.
The fact that only one quadrature is encoded has an important implication, in the sense that the squeezed EPR correlations in Eq.~\eqref{eq:AliceBobPostChannel} in the amplitude quadrature are not accessible to Alice and Bob.
In other words, they have no way of estimating $T_Q$.
$T_P$ is estimated as normal, and the variance of the amplitude quadrature has to be monitored to detect the level of excess noise in this quadrature.
However, monitoring the variance alone is not enough to determine $T_Q$.
Fortunately, it is possible to place a bound on the possible values of $T_Q$, through Heisenberg's inequality.
In the covariance matrix description the inequality has the form $\gmat{\Gamma_{A B \kappa}} + i \gmat{\Omega} \geq 0$, i.e.\ the matrix $\gmat{\Gamma_{A B \kappa}} + i \gmat{\Omega}$ must be positive semi-definite. 
The physicality is tested by a simple numerical determination of the eigenvalues of this matrix, as they must be non-negative for $\gmat{\Gamma_{A B \kappa}}$ to be physical.
With this bound one can map out a region of security in $(T_P, T_Q)$ space.
This is done in Fig.~\ref{fig:TPTQgrid} where excess noise is added, and from Fig.~\ref{fig:TPTQgrid}(a) to Fig.~\ref{fig:TPTQgrid}(b) there is a clear increase in the size of the region of physicality.
If neither excess nor preparation noise is present, the bound enforces the relation $T_P = T_Q$.
However, even minute amounts of noise will widen the bound and since $T_Q$ is controllable by the eavesdropper a worst case scenario approach is taken and $T_Q$ is thus always chosen to give the lowest possible rate.

\begin{figure}[ht]
    \begin{center}
    \subfloat{\includegraphics[width=0.49\textwidth]{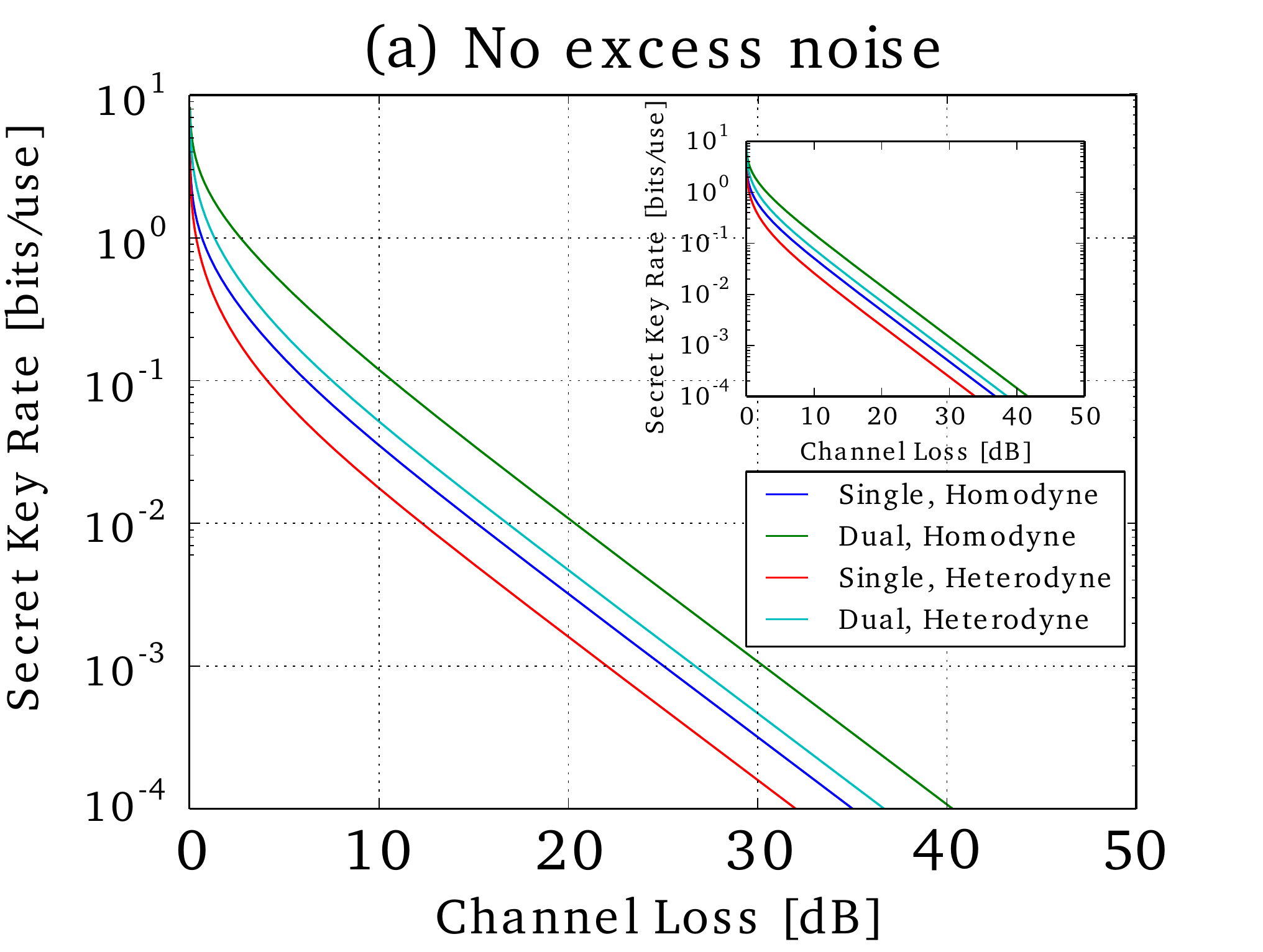}\label{fig:NoNoiseTheoryRR}} 
    ~
    \subfloat{\includegraphics[width=0.49\textwidth]{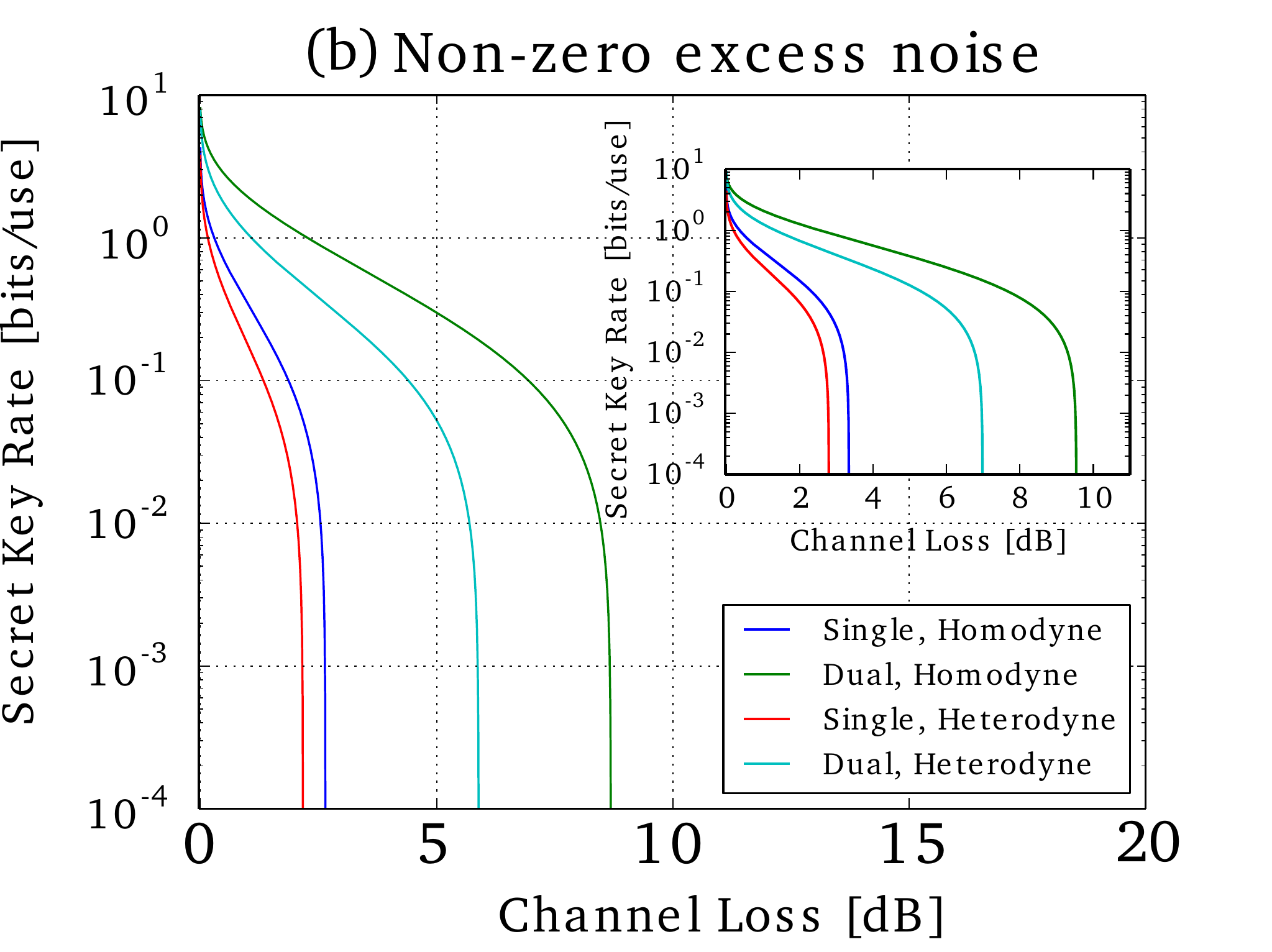}\label{fig:NoiseTheoryRR}}
    \end{center}
\caption{Theory plots of secret key rates for protocols using single- and dual-quadrature modulation respectively, with homodyne and heterodyne detection strategies. Both plots use reverse reconciliation, and plot (a) has no excess noise, while for (b) $W_Q = W_P = 1.05$ SNU. The modulation variance was optimized for each protocol and channel attenuation in the $P$ quadrature to obtain the largest possible secret key rate. $\beta = 97\,\%$ except for both insets where $\beta = 100\,\%$ was assumed.}
    \label{fig:RRTheory}
\end{figure}

In Fig.~\ref{fig:RRTheory} we plot the secret key rate as a function of channel attenuation for different detection strategies both for the 2D and the 1D Gaussian modulation patterns, using reverse reconciliation.
In Fig.~\ref{fig:RRTheory}(a) there is no excess noise, while in Fig.~\ref{fig:RRTheory}(b) $W_Q = W_P = 1.05$ SNU. In both cases we assume no preparation noise. The modulation variance was optimized for each channel attentuation value to obtain the largest possible secret key rate.
The insets of Fig.~\ref{fig:RRTheory} shows the secret key rates for an ideal error reconciliation efficiency of $\beta = 100\,\%$, while a realistic value of $\beta = 97\,\%$ was assumed for the main figure.

\begin{figure}[ht]
    \begin{center}
    \subfloat{\includegraphics[width=0.49\textwidth]{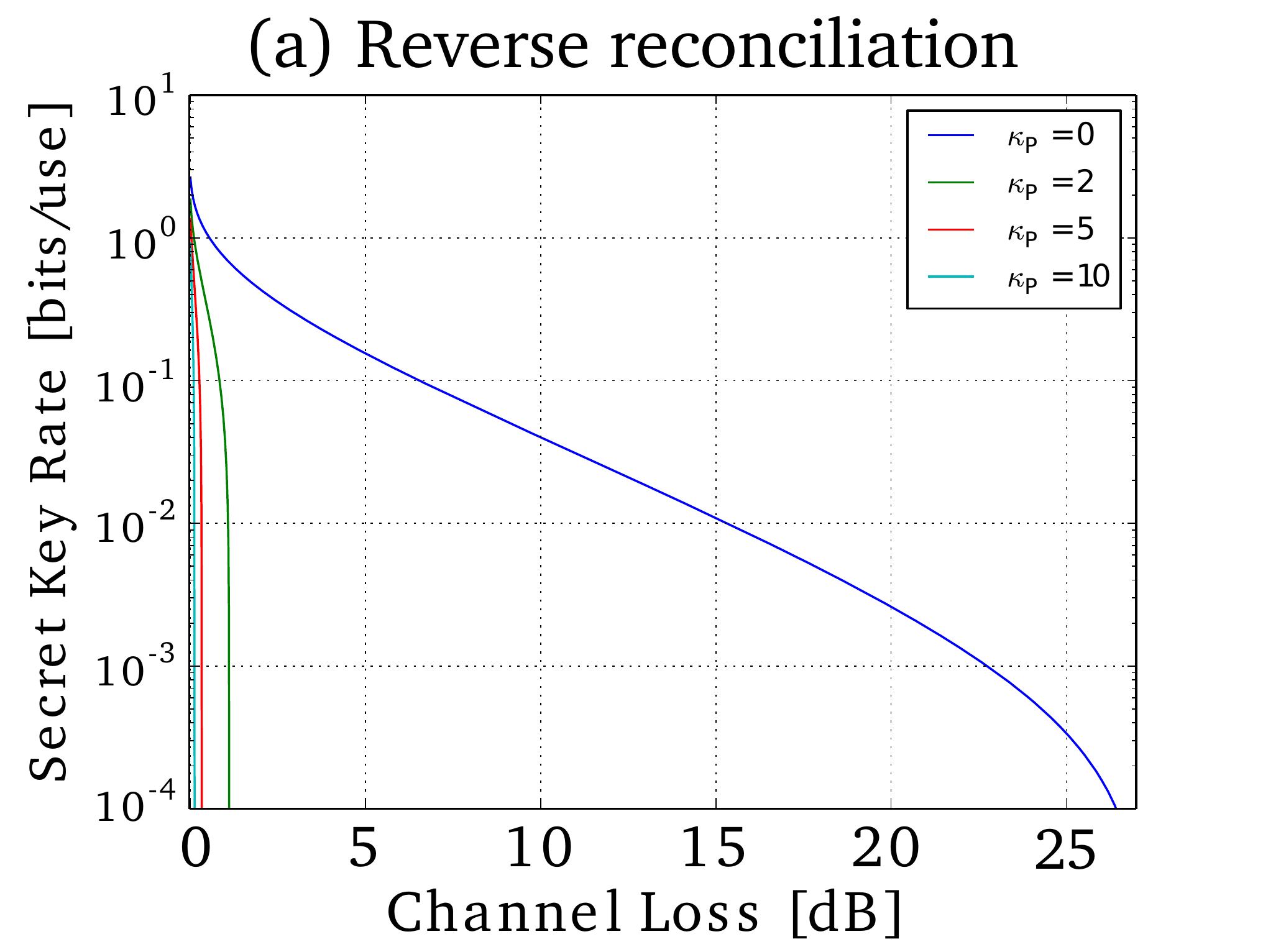}\label{fig:kappaRatesRR}} 
    ~
    \subfloat{\includegraphics[width=0.49\textwidth]{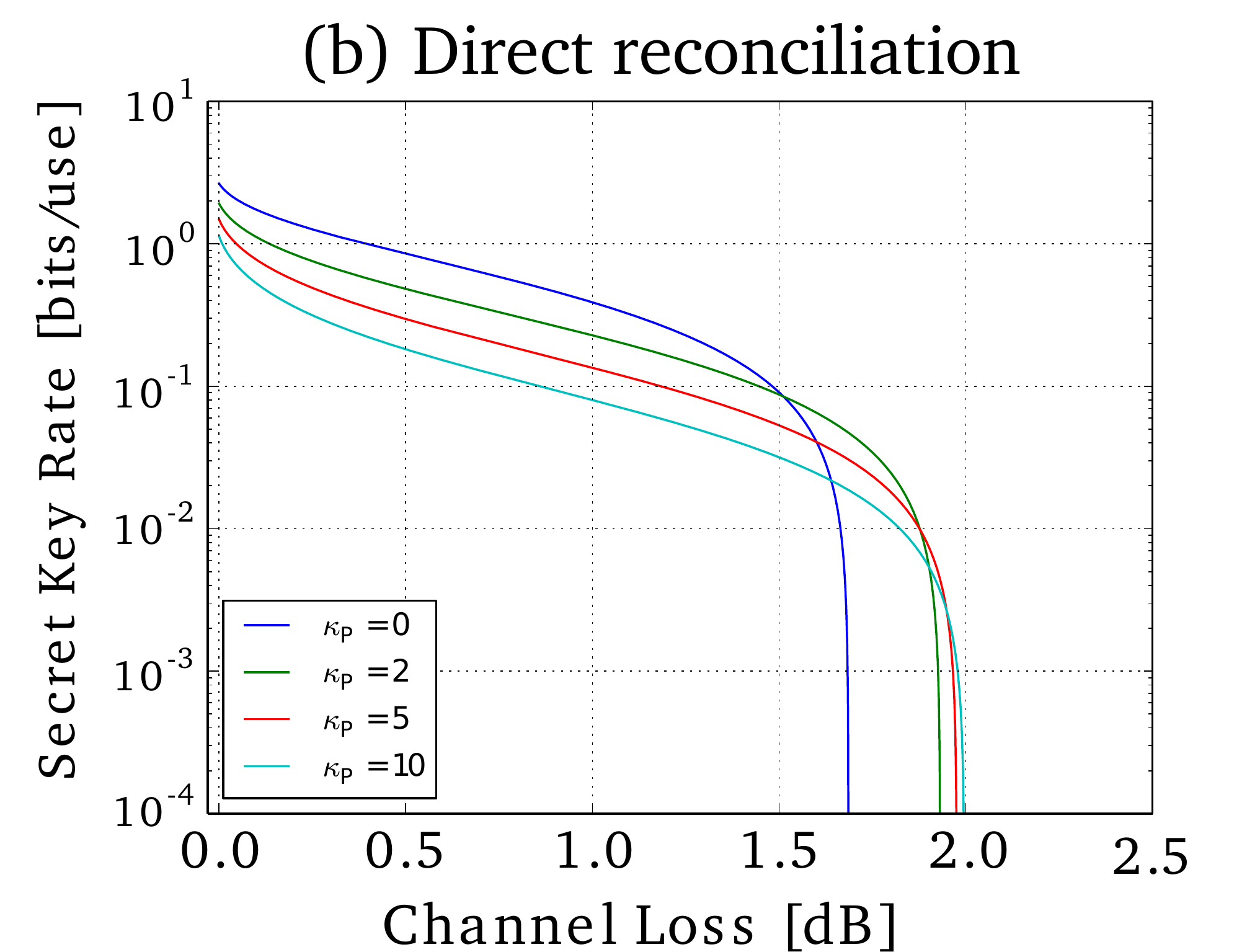}\label{fig:kappaRatesDR}}
    \end{center}
\caption{Theory plots of secret key rates with varying degrees of preparation noise in the encoded quadrature. Plot (a) uses reverse reconciliation, while for (b) direct reconciliation is used. $\beta = 1$, $\mu = 1000$ and $W_Q = W_P = 1$.}
    \label{fig:kappaTheory}
\end{figure}

In Fig.~\ref{fig:kappaTheory}, the secure key rate is plotted for different levels of preparation noise, with reverse reconciliation in Fig.~\ref{fig:kappaTheory}(a) and direct reconciliation in Fig.~\ref{fig:kappaTheory}(b). 

As was reported by Weedbrook \textit{et al.}, in \cite{Weedbrook2012a} and Jacobsen \textit{et al.}, in \cite{Jacobsen2015}, the single quadrature protocol retains its robustness to preparation noise when using direct reconciliation. This is shown in Fig. \ref{fig:kappaTheory}.
From this it is also evident that an optimal non-zero level of preparation noise exists.

From the plots in Figs. \ref{fig:TPTQgrid}, \ref{fig:NoiseTheoryRR} and \ref{fig:kappaTheory}, it is clear that a single modulated quadrature combined with the right channel offers quantum security, albeit slightly reduced from a protocol using both quadratures.
In general one has to trade lower key rates and a reduction in tolerated channel loss for experimental simplicity.
In the single quadrature scheme, half of the measurements ($Q$-quadrature measurements) are not used for key generation but only for state estimation. This naturally reduces the rate by a factor of two.
It can, however, be partially compensated by introducing an asymmetry in the heterodyne detector or the switching probability of the homodyne detector.

\section{Experiments}

\begin{figure}[ht]
    \centering
    \includegraphics[width=8.5cm]{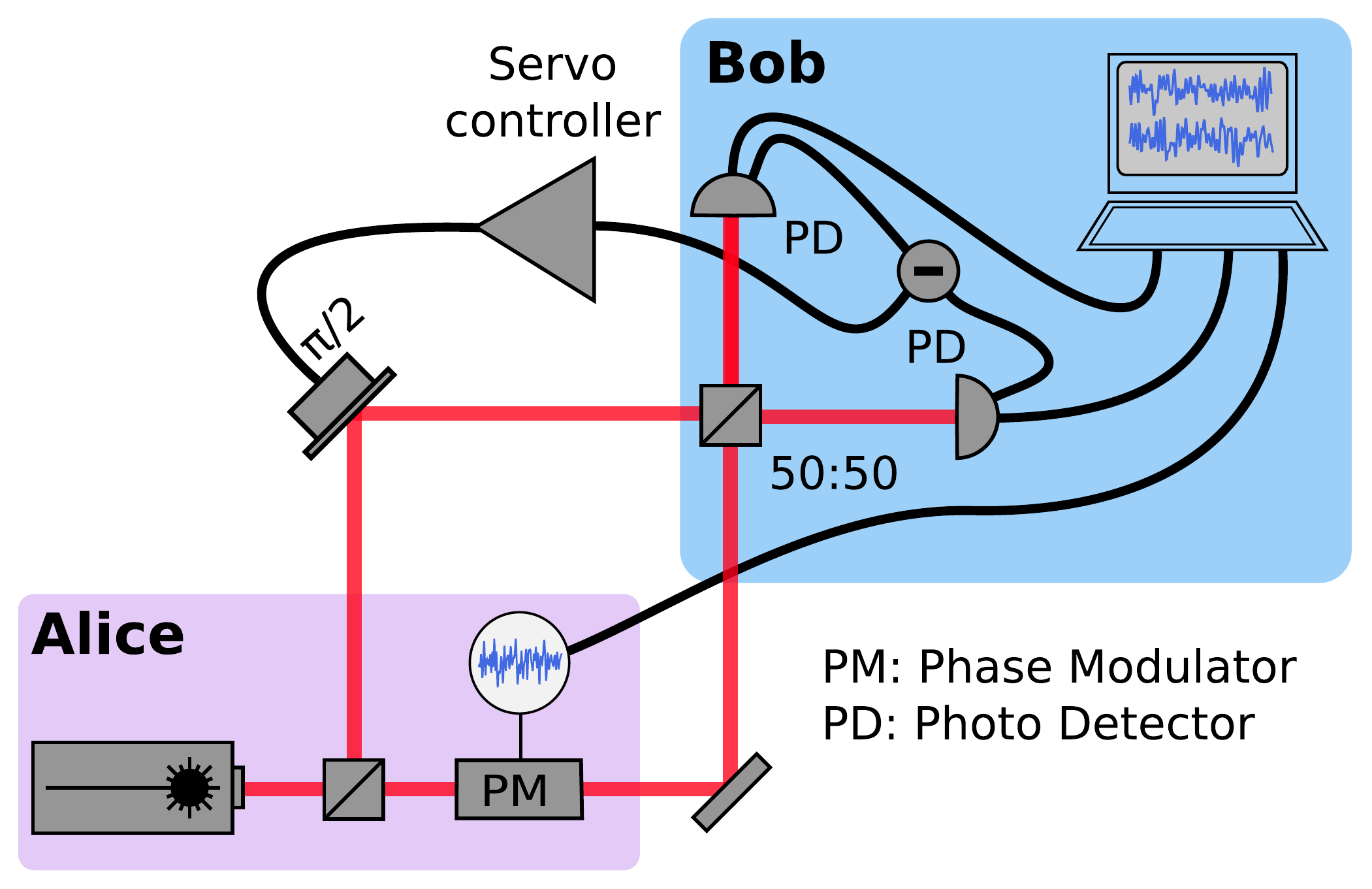}
    \caption{Schematic of the experimental setup. At Alice's station a laser beam was split into a reference and a signal beam. The signal beam was phase modulated with Gaussian white noise. After transmission of both beams to Bob, he performed heterodyne detection by interfering the two equally bright beams at a balanced beam splitter while the phase was locked to $\pi/2$. Homodyne detection was done by brightening the local oscillator while dimming the signal beam to satisfy the brightness approximation. The two outputs of the two photo detectors as well as the noise from the Gaussian white noise generator were recorded.}
    \label{fig:experiment}
\end{figure}

We demonstrate the single-quadrature CVQKD schemes with heterodyne and homodyne detection using bulk optical components. A schematic of the experimental setup is shown in Fig.~\ref{fig:experiment}. 
The optical power and the electronics can be adjusted to make the detection stage function as either a heterodyne or homodyne detector.

For heterodyning a $1064$\,nm continuous-wave laser beam was split equally into a reference (local oscillator) and a signal beam, each carrying a power of $2.8$\,mW. For homodyning, the signal power was adjusted to $0.1$\,mW, while the reference beam (local oscillator) was kept at $2.8$\,mW. The signal beam was modulated in phase with an electro-optical phase modulator using Gaussian white noise from a white noise generator.
The reference as well as the signal beam were then transmitted to Bob who performed heterodyne detection which was implemented by interfering both beams at a balanced beam splitter and locking the relative phase to $\pi/2$. For homodyne detection were interfered at the same beam splitter, but the $P$ and $Q$ quadratures were measured by switching the quadrature between runs, by locking the relative phase between the signal and reference beams to $\pi/2$ and $0$ respectively.
The AC outputs of the photo detectors were demodulated at $10.5$\,MHz and low-pass filtered at $100$\,kHz before being sampled with a $14$\,bit data acquisition card with a sampling rate of $500$\,kHz.
For heterodyning, in post-processing the sum and difference of the two sampled data streams were calculated. 
These outputs represent measurements of the amplitude and phase quadrature amplitude, respectively, when both beams have the same optical power.
In the homodyne measurement, $Q$ and $P$ quadrature data is found in post-processing by the difference of the two sampled data streams.
In addition to these measured data (at Bob), we also recorded the data resulting from the white noise generator (at Alice).

The vacuum reference was measured by disconnecting the white noise generator from the phase modulator.
After that the phase modulation was first set to a modulation variance of $15$\,dB above vacuum which set the $100\,\%$ transmission value.
For heterodyning, the modulation variance was subsequently reduced to simulate optical loss in the channel. Since only coherent states were involved in our implementation this procedure is equivalent to introducing optical loss. However, this scheme enabled us to perform heterodyne detection in the form described above, since the requirement of having the same optical power in reference and signal beam was fulfilled. For homodyne detection, loss was introduced via beamsplitters and waveplates, as there is no power requirement in the signal beam in this configuration.

For each measurement run we recorded $10^6$ samples and estimated the excess noise and transmission of the channel. The magnitude of the correlations directly defines the level of the mutual information, through the formula
\begin{equation}
I(A:B) = \dfrac{1}{2} \log_2 \left(\dfrac{V_B}{V_B - C_{AB}^2/V_A} \right)\ ,
\end{equation}
where $V_B$ is the variance of Bob's measurement in the signal quadrature, $V_A$ is the variance of Alice's input signal, and $C_{AB}$ is the covariance between the data sets.
The preparation noise was determined at $100\,\%$ transmission where no excess noise introduced by the eavesdropper is present.
This noise level is calculated by subtracting the (properly scaled) input with the output.
In the limit of perfect correlations, one would expect such a subtraction to yield zero.
Any residual variance is therefore to be regarded as noise.
If the correlations are poor, the preparation noise consequently goes up.
For transmissions lower than unity, the combined excess and preparation noise is again determined by subtraction.
The scaling of the preparation noise with transmission is known, so any leftover noise in the combined noise must be ascribed to the eavesdropper, i.e.\ excess noise.

\subsection{Heterodyne results}

\begin{figure}[ht]
    \centering
    \includegraphics[width=8.5cm]{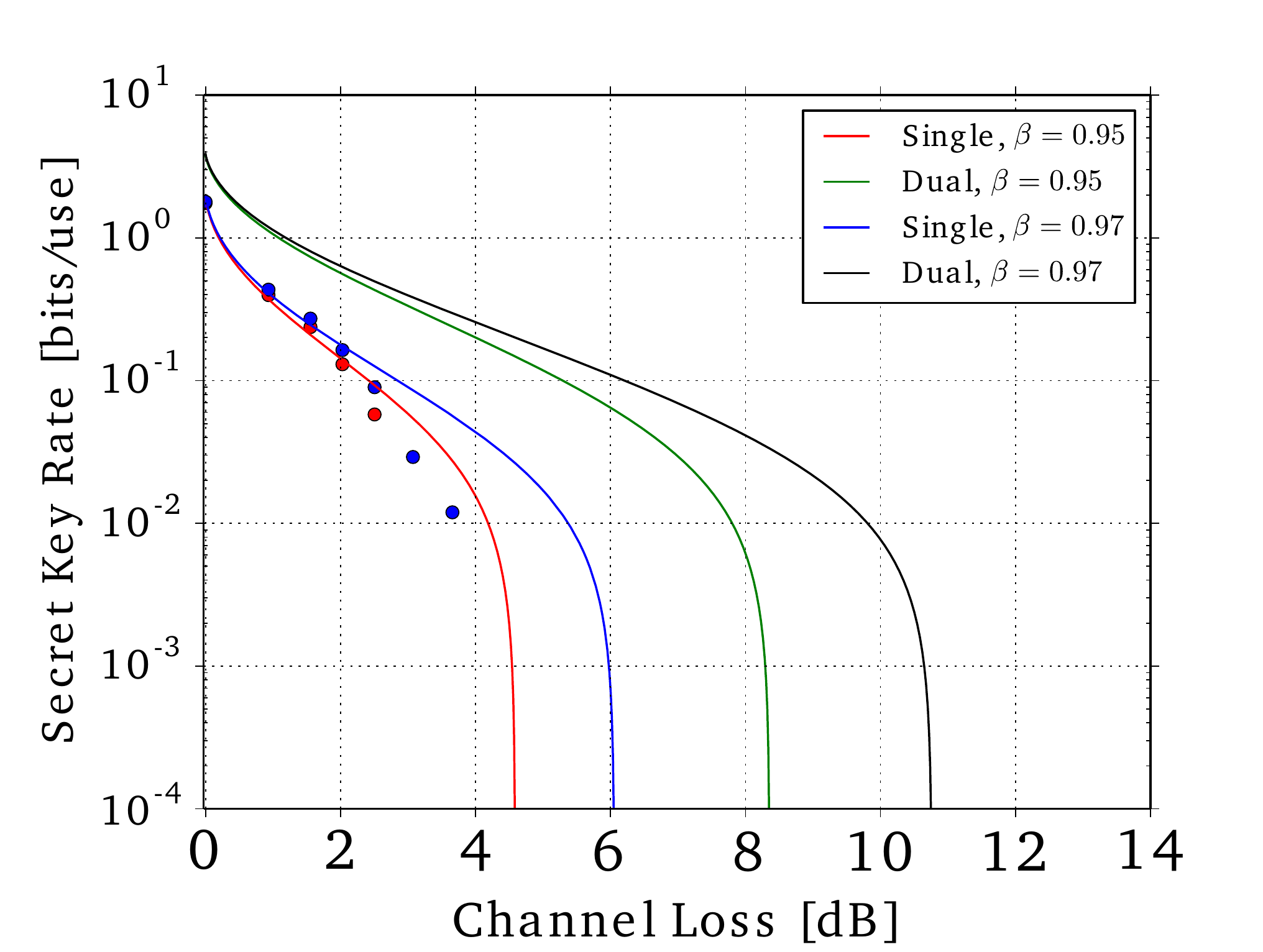}
    \caption{Experimental results showing the secret key rate calculated from the measured data for an error reconciliation efficiency of $\beta = 95\,\%$ and $97\,\%$, respectively, with heterodyne detection and reverse reconciliation. The red and the blue solid lines denote theory curves calculated from the channel parameters. Error bars are smaller than the point size. The theoretical model includes excess noise, which on average is $W_Q = 1.004$ SNU and $W_P = 1.005$ SNU, respectively, and preparation noise, which is transmission dependent trusted noise, an artifact of the imperfect modulation. It is $\kappa_Q = 0.07$ SNU and $\kappa_P = 0.025$ SNU, and $\mu = 31.2$ SNU. For comparison we show the secret key rates that could be obtained using dual-quadrature modulation with heterodyne detection, but otherwise the same channel.}
    \label{fig:HeterodyneResults}
\end{figure}

The results for heterodyne detection are shown in Fig.~\ref{fig:HeterodyneResults}.
For heterodyne detection the preparation noise was determined to $\kappa_Q = 0.07$ SNU and $\kappa_P = 0.025$, which is related to imperfect modulation performed by Alice due to a mismatch between the electro-optical modulation and the recorded data stream of the from an analogue white noise generator.
The asymmetric excess noise levels $W_Q = 1.004$ SNU and $W_P = 1.005$ SNU, respectively, were on the other hand rather small.
The red and blue solid lines in the figure are theory curves calculated with the above parameters and show good agreement with the measurement data.
The extrapolated maximum transmission line for the single-quadrature modulation with heterodyne detection is slightly above $20$\,km for an error reconciliation efficiency of $\beta = 95\,\%$ if an optical fiber with $0.2$\,dB/km attenuation is employed.
For an error reconciliation efficiency of $\beta = 97\,\%$ about $30$\,km is achievable.
Note that the modulation variance in our experiment was fixed and not optimized as in the theory plots in Fig.~\ref{fig:NoiseTheoryRR} since we wanted to show the agreement of measurement and theory with a simple dependence on the channel loss. 
For comparison we added theory curves for a dual-quadrature modulation scheme (green and black solid lines) with the same channel parameters, but twice the alphabet size ($\mu - 1$ signal in both quadratures).
In the intermediate region of $25$\,km or equivalently $5$\,dB channel loss, the secret key rate of our single-quadrature modulation scheme is merely a factor of about $10$ lower than with two modulations in this particular parameter space ($\beta = 97 \, \%$).

\subsection{Homodyne results}

\begin{figure}[ht]
    \begin{center}
    \subfloat{\includegraphics[width=0.49\textwidth]{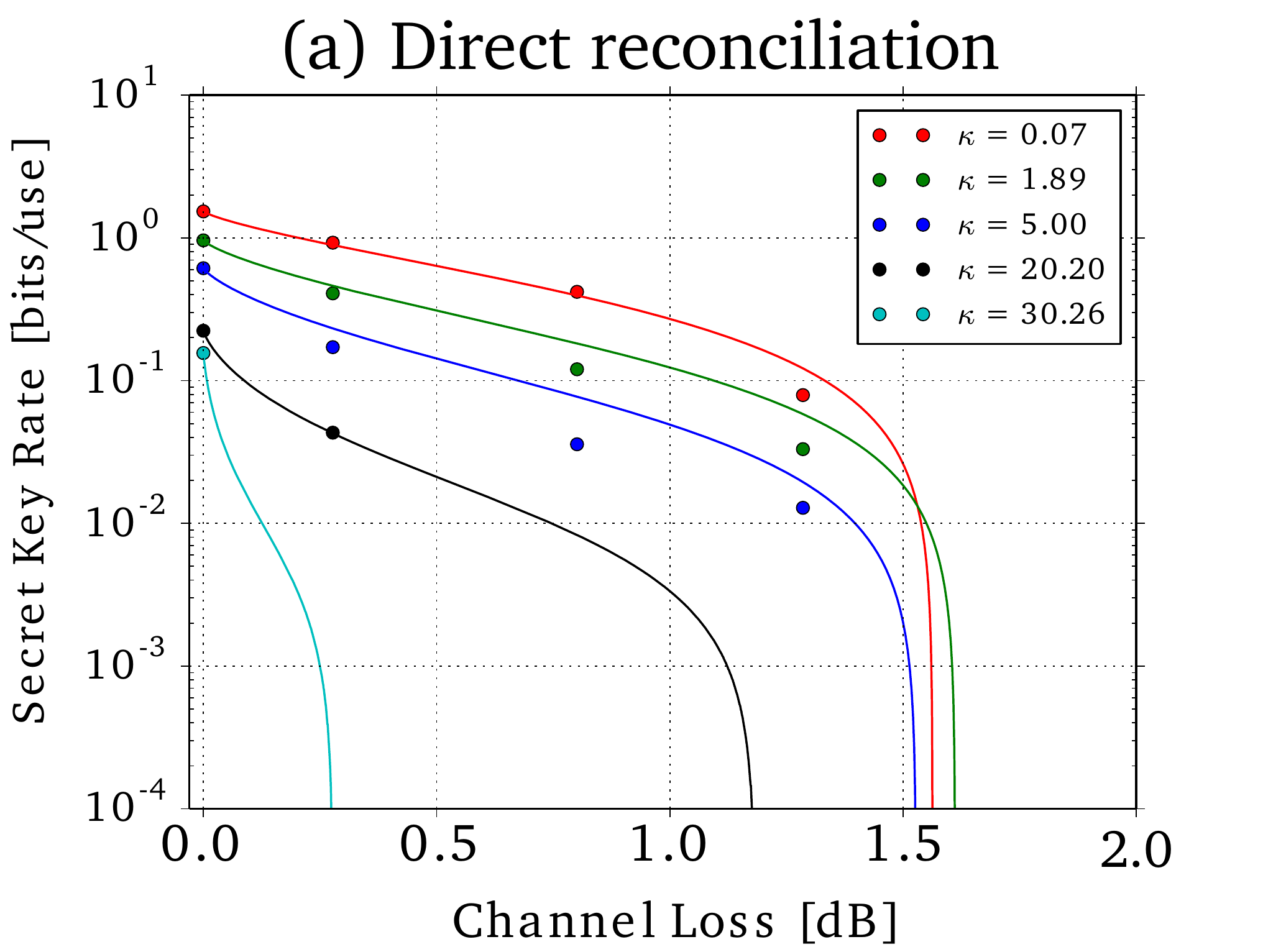}\label{fig:HomodyneResultsDR}} 
    ~
    \subfloat{\includegraphics[width=0.49\textwidth]{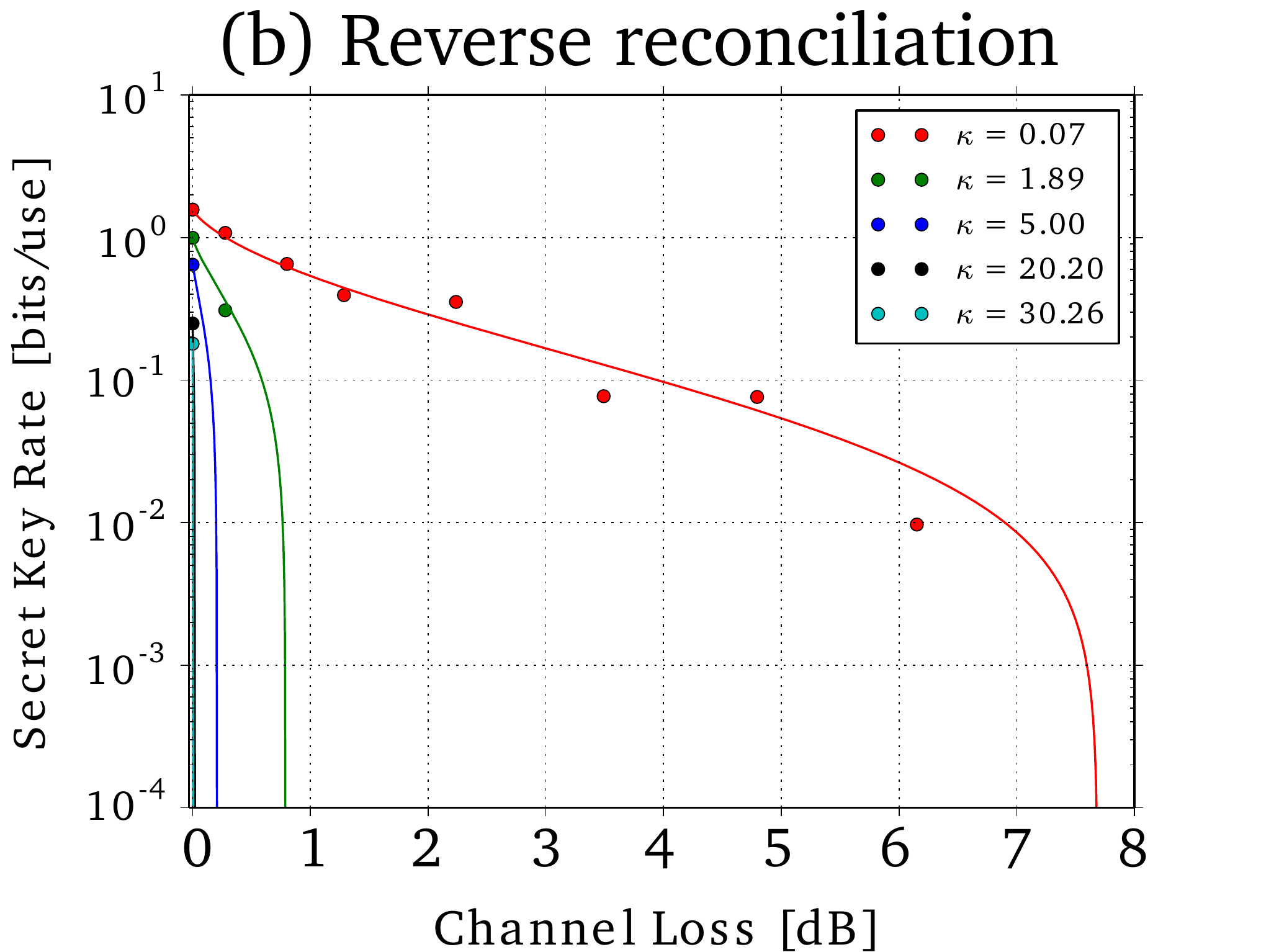}\label{fig:HomodyneResultsRR}}
    \end{center}
\caption{Experimental results showing the secret key rate calculated from the measured data for an error reconciliation efficiency of $97\,\%$, with homodyne detection and (a) direct reconciliation or (b) reverse reconciliation. The multiple coloured solid lines denote theory curves calculated from the channel parameters, with varying degrees of $P$-quadrature preparation noise. Error bars are smaller than the point size. The theoretical model includes excess noise, which on average was $W_Q = 1.01$ SNU. Due to imperfections $W_P$ scaled with the preparation noise to a maximum value of $W_P = 1.88$ SNU at $\kappa_P = 30$ SNU. $\mu = 31$ SNU, and in the unencoded quadrature $\kappa_Q = 0.03$ SNU. Direct reconciliation shows robustness against the preparation noise as expected, though the detrimental scaling of the excess noise diminishes the positive effect.}
    \label{fig:HomodyneResults}
\end{figure}

The results for homodyne detection are shown in Fig.~\ref{fig:HomodyneResults}.
For homodyne detection the preparation noise was determined to $\kappa_Q = 0.03$ SNU.
The asymmetric excess noise was $W_Q = 1.01$ SNU. Due to experimental imperfections in the modulators $W_P$ scaled with the preparation noise to a maximum value of $W_P = 1.88$ SNU at $\kappa_P = 30$ SNU.
The high modulation variance of the preparation noise in the signal quadrature is responsible for the increase in $P$ quadrature excess noise, since the modulators are fed with larger signals.
The coloured solid lines represent theoretical rates calculated from the channel parameters, and show good agreement with the measurement data, though it is not surprising to see dramatic reductions in distance considering the high excess noise.
An extrapolation similar to the one performed for the heterodyne data gives a maximal distance of about $20$\,km for an error reconciliation efficiency of $\beta = 97\,\%$.

\section{Conclusion}

In conclusion, we have investigated a new, simplified CVQKD scheme based on single-quadrature modulation as opposed to the traditional dual-quadrature modulation scheme. In contrast to DVQKD, modulation in conjugate bases is not a requirement for secret key generation in CVQKD.
The secrecy stems from the non-orthogonality of different coherent states which is obtainable in a single quadrature basis.
In this paper we have proven the security for such a scheme against collective attacks in the asymptotic limit, and we have demonstrated the protocol experimentally for both heterodyne and homodyne detection, and included preparation noise in the theoretical description.
Due to the extraordinary simplicity of the scheme, we expect that it will gain commercial interest. 

\section*{Acknowledgments}
This research was supported by the Danish Agency for Science, Technology and Innovation (Sapere Aude) (0602-01686B). T.G thanks the H.C. \O rsted postdoc programme for support.

The authors would to thank Vladyslav Usenko, Frederic Grosshans, Torsten Franz and Anthony Leverrier for enlightening discussions regarding the security of the scheme.

\end{document}